# Vibration diagnostics instrumentation for ILC

**A Bertolini**
DESY, Notkestrasse 85, 22603 Hamburg, Germany

E-mail: alessandro.bertolini@desy.de

**Abstract.** The future $e^-e^+$ 500 GeV International Linear Collider will rely on unprecedented nanometer scale particle beam size at the interaction point, in order to achieve the design luminosity. Tight tolerances on static and dynamic alignment of the accelerator cavities and optical components are demanded to transport and focus the high energy electron and positron beams with reasonable position jitter and low emittance. A brief review of techniques and devices evaluated and developed so far for the vibration diagnostics of the machine is presented in this paper

## 1. Introduction

The proposed electron-positron International Linear Collider (ILC) is being designed to operate with an initial 500 GeV center-of-mass energy (an upgrade to 1 TeV is foreseen). Low emittance, 40 nm·rad in the vertical, and unprecedented beam size at the interaction-point (IP), 5.7 nm the vertical and 639 nm the horizontal, are required to achieve the goal of $2\times10^{34}\text{cm}^2\text{sec}^{-1}$ peak luminosity. Even though the choice of L-band superconducting technology for the ILC has relaxed significantly the tolerances on static and dynamic alignment of cavities and magnets, transporting the $e^-e^+$ beams, from the damping rings to the beam delivery system (BDS), through the more than 14 km long linac sections, without degrading the nominal emittance, remains challenging. In both main linac and BDS sections of the machine, after the initial static and beam-based alignment, pulse-to-pulse local feedback systems, using a few corrector magnets (or stripline kickers) and the readout from few selected beam position monitors (BPM), will control the beam orbit in the 0.1-1 Hz frequency range. Bunch-to-bunch (intra-train) feedback, allowed by the 5 Hz pulse repetition rate with a long train of largely spaced bunches (~3000 bunches in ~1 msec), could be used at critical locations to correct faster dynamic misalignment, whose dominant source will be the uncorrelated mechanical vibration of quadrupole magnets due to ground motion and mechanical systems. The intra-train feedback at the IP will compensate for vibration-induced jitter on the final-focus magnets, by steering the electron and positron into collision. The combination of beam-based feedback systems and siting of the ILC in a low noise seismic environment (particularly having low cultural noise content, usually dominating above 1 Hz), is envisaged to mitigate these effects, which could degrade the collider luminosity. This scenario, depicted in the ILC Reference Design Report (RDR) [1], reserves to mechanical vibration measurements a role of diagnostics, which will be particularly important for the design and validation of the components, during the commissioning of the machine and to improve feedback convergence time. A brief review of the technologies which have been studied and evaluated for the different sections of the machine, is presented.

## 2. Linacs

In the linacs fast beam position jitter can build up as the incoherent superposition of the fast motion of the 560 quadrupoles. Uncorrelated vertical vibration level of the order of 30 nm root-mean-square (rms), as expected for a typical quiet site [2], is well tolerable. A level of 100 nm rms still lead to a negligible pules-to-pulse emittance growth at the end of the linac, but the resulting oscillation in the vertical plane (one to two times the vertical beam size) in the BDS could lead to significant emittance dilution from sources such as collimator wakefields. In this case an intra-train feedback at the exit of the linac could be necessary. While the ground motion at all of the sites complies with these tolerances, care must be taken in vibration isolation from in-tunnel and near-tunnel hardware, and in the mechanical design of the cryomodule components (low frequency mechanical resonances can largely amplify the ground motion beyond tolerable levels).

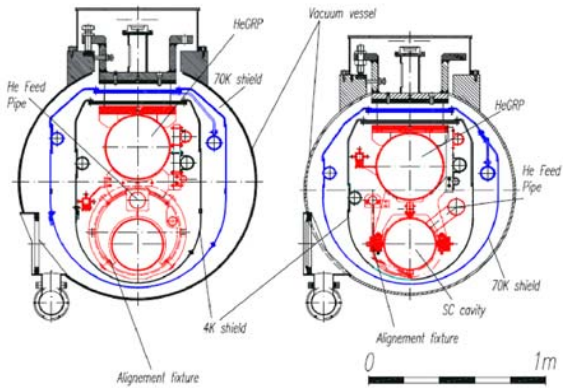

**Figure 1.** Design evolution of TTF cryomodules from second generation, on the left, to the third generation, on the right. The helium gas return pipe (GRP) is the main structural element supporting both the cavity string and the quadrupole package [3].

ILC cryomodules (so called type IV cryomodule) will be a further evolution of the third generation type (so called type III cryomodule [3]) used for the module 4 and 5 (and module 6 in the next future) in the Tesla Test Facility (TTF) linac, at DESY (see the cross section of a TTF Type-III cryomodule in Figure 1 right side).

Nevertheless, except for the proposed new location of the quadrupole package at the center of the module, the main features and the layout of the components inside the cryogenic vessel, with the helium gas return pipe (GRP) as the main structural element, will be preserved.

Recent studies, performed at room temperature [4-5] on a second generation TTF cryomodule (see the cross section in Figure 1), have proven the reliability of the conceptual design, showing no resonances from the quadrupole/cavity string support structure below 20 Hz in the horizontal and below 40 Hz in the vertical direction.

These results are of relevant interest and promising for the ILC, particularly considering that in the improved Type-III cryomodule version, stiffer support for the quadrupole package position have been introduced and GRP posts and vacuum vessel supports have been moved to reduce the sensitivity to misalignment and external forces during insulation vacuum pumping and cooldown.

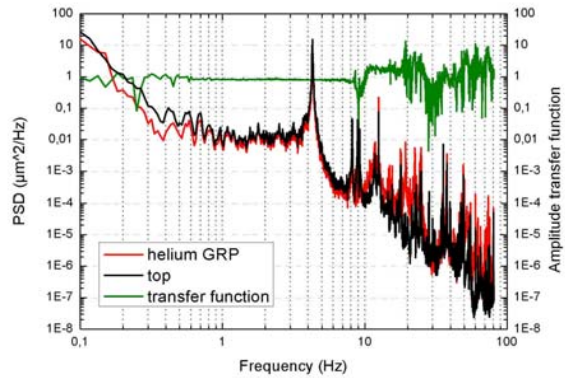

**Figure 2.** Horizontal transfer function from the vessel top to the GRP, measured with two seismometers on a TTF Type-II cryomodule. The measurement was done with the module disconnected from the linac. The displacement power spectra are dominated at low frequency by the rigid body modes of the cryostat on its support system and, at high frequency by near field vibration sources in the experimental area. The amplitude transfer function shows no evidence for mechanical resonances below 20 Hz [4-5].

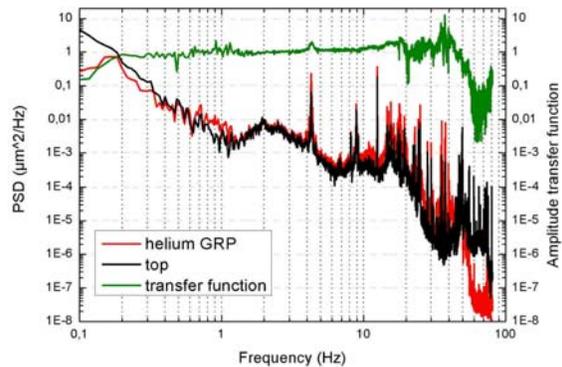

**Figure 3.** Vertical transfer function from the vessel top to the GRP. Same setup than in Figure 2. No resonant structures are visible in the amplitude transfer function below 40 Hz [4-5].

The same experiments also have shown that the cold mass motion is dominated at low frequency by the resonances due to the rigid body modes of the cryostat on its support system (see Figures 2-3). Improvement is mandatory in this direction, while further investigation of possible noise from the refrigeration system is envisaged in the next future. Nanometer level vibration diagnostics inside a cryomodule has been initially pursued by means of standard inertial sensors (piezoelectric accelerometers), mounted aboard the quadrupole package. Piezoelectric accelerometers are limited in low level, low frequency measurements by the electromechanical efficiency of the piezoelectric material used (charge sensitivity) and by the characteristics of the preamplifier.

At low frequency, the minimum detectable acceleration has the typical behaviour of ~1/f and it is inversely proportional to charge sensitivity [6]. State-of-the-art piezoelectric sensors [7-8] have spectral sensitivities better than $10\,\text{nm}/\sqrt{\text{Hz}}$ at 1 Hz, achieved with low resonant frequency mechanical design (less than 1 kHz) and built-in low-noise FET preamplifier. These devices are not suitable for cryogenic operation and sensors with remote charge amplifier must be used. The best charge-mode accelerometer available off-the-shelf (Model 7703A-1000 from Endevco, Co.) is currently installed in TTF cryomodules, aboard the quadrupole helium vessel. Unfortunately, the loss of a factor of ~3 in charge sensitivity, when cooled down to 4.5°K, further lowers their relatively poor resolution, finally lying in the µg's region between 1-10 Hz [9]. For these reasons, these sensors look marginally or not sufficiently sensitive to detect vibrations levels of interest for linac diagnostics in that frequency band. Anyway, at higher frequencies, they look suitable to monitor possible excess acoustic noise introduced by the refrigeration system or mechanical vibrations produced by near-field sources. Better resolution in the low frequency band could be achieved by using geophones, and tests are foreseen in DESY. Geophones have low output impedance, work with remote low-cost signal conditioning electronics and have been demonstrated to be compatible with a cryogenic environment [10]. In their basic arrangement, these sensors provide flat inertial velocity output above the fundamental resonant frequency of their mass-spring system. Anyway their response can be easily extended to lower frequencies by using inverse filtering or overdamping schemes in the readout amplifier [11]. For example, the response of the industry-use classic 4.5-Hz geophone (having size comparable with the piezo accelerometers presently used in TTF linac) can be extended down to 0.25 Hz preserving a self-noise around $1\,\text{nm}/\sqrt{\text{Hz}}$ at 1 Hz [12].

An alternative choice for the diagnostics of cryomodules is to use displacement sensors (RF bridges, interferometers, laser Doppler velocimeters etc.) measuring the relative position of the quadrupole with respect to the cryostat. In this case a complementary inertial sensor mounted outside on the cryostat vessel is desirable, in order to take into account the effect of the common mode motion.

In present design, TTF cryomodules are equipped with a Wire Position Monitor (WPM) system. Designed to monitor with high precision displacements and misalignments between the cold mass and the cryostat, the WPM has been recently proposed as mechanical vibration sensor [13]. The horizontal and vertical position of a copper-beryllium stretched wire with respect to a reference bore, rigidly connected to the helium-gas return pipe, is measured at different positions with a bridge-like RF sensor. Seven detectors placed in critical positions (at each end, at and between the three posts supporting the cold mass) are used in each module. Each detector works as a sort of microstrip four channel directional coupler. A 140 MHz RF signal is applied to the stretched wire, nominally placed at the center of the monitor bore and is sampled by four microstrips, leading to the wire relative position in both the horizontal and vertical planes. Vibration detection is performed by a super-heterodyne receiver where the base-band demodulation is made digitally.

Minimum detectable relative displacements around $100\,\text{nm}/\sqrt{\text{Hz}}$ at 1 Hz have been proven, over a 74 dB dynamic range. Self-oscillations of the wire limit the mechanical measurement range and limit the linear response bandwidth to a fraction of the frequency of the first violin mode (about 6 Hz in the present design). The relatively poor resolution and the strong non-linearity of the response above a few Hz make the WPM, at the moment, suitable only for a qualitative diagnostic. High resolution over larger bandwidths is achievable by optical position sensors, even if they appear suitable only for design studies and tests, because of the relative complexity and costs. The extreme environmental conditions and the huge longitudinal shrinking of structures during cool-down put severe constraints on the sensor design, which must use target mirrors or retroreflectors attached to the surface of the cold mass. The target position with respect to a reference mirror, placed on the outer wall of the cryostat, can then be detected with interferometric techniques.

An alternative layout, based on a 2x2 fiber coupler, with all in-vacuum optics was successfully used for testing stability of LHC cryodipoles during thermal cycles [14], and it could be reconsidered.

## 3. Beam delivery system and final focus

Beam size reduction factors of a few hundreds are foreseen in the beam transport system from the exit of the linac to the IP. The strong magnet strength of the demagnifying optics and the corresponding large beta functions lead to the tightest tolerances on field quality and alignment along the machine.

Particularly at the final doublet (FD) the correspondence between a transverse displacement of the quadrupoles and the offset of the beam at the IP is one-to-one. Besides beam-based orbit feedback in the BDS, a fast intratrain beam-beam position feedback will be implemented at the IP, to correct for vertical and horizontal position offsets at the collision point [15]. The IP fast-feedback system relies on the strong beam-beam interaction dynamics for its operation. For nm-scale offsets at the IP, beam-beam deflections from 10 to hundreds of µrad are produced. Then BPM with a modest 10 µm resolution allows to control the IP beam position to the ~$0.1\sigma$ level ($\sigma_y$ = 5.7 nm) level, roughly controlling the luminosity to better than the 2%. The capturing range of the feedback system, from 30 to 170 nm depending on the chosen ILC parameter set, is also very large. These specs are the reason why no active means of mechanical stabilization has been envisaged in the baseline configuration of the final doublet.

Anyway, vibration measurement of FD is part of the baseline, to give diagnostics for the fast beam-beam feedback itself. Foreseen implementations include accelerometer on cryostat, interferometry from cold mass to cryostat and interferometry from the cold mass to an external reference (the preferred one).

The final doublet will be located inside the yoke of the detector. Inertial sensors must fit the tight space available and withstand with the strong solenoidal field (a few teslas) and the high radiation environment.

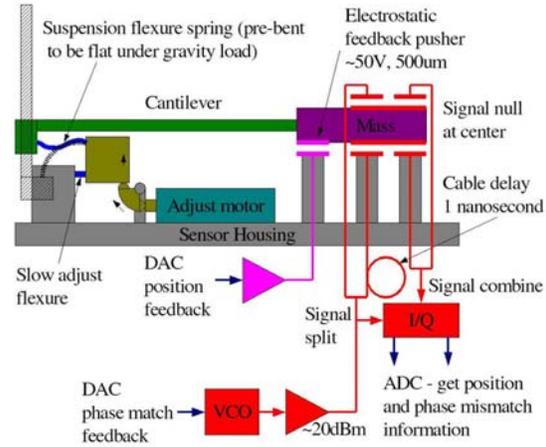

**Figure 4.** Block scheme of the NLC vertical accelerometer: The motion of the proof mass with respect to the instrument's frame is read by an RF capacitance bridge. An electrostatic actuator provides the force necessary to dynamically keep the mass at the equilibrium position. The feedback algorithm is digitally implemented by using a DSP.

Standard seismometers, based on low resonant frequency (0.25-1 Hz) pendulum, are by far the most sensitive instruments available. Unfortunately, the pendulum's elastic elements are realized with metallic alloys, selected for their low thermoelastic coefficient, which are ferromagnetic (of the family of NISPAN-C), thus making the instruments highly sensitive to external magnetic fields and unsuitable for the IP region. A novel inertial sensor, capable to operate in such an environment, was prototyped by NLC community in SLAC in the past few years [16]. The instrument was built with non-magnetic material in all the parts. A 40-gr tungsten mass was suspended with a pre-bent cantilever spring made of copper-beryllium, resulting in a vertical resonant frequency of about 1.5 Hz.

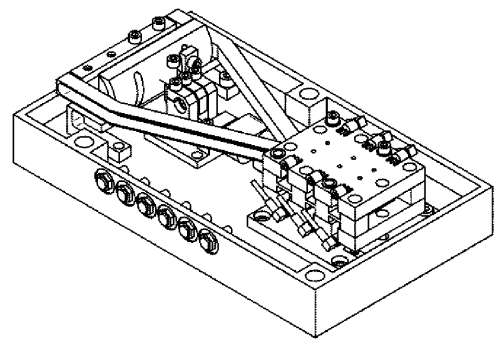

**Figure 5.** Full view of the prototype of the non-magnetic inertial sensor developed by NLC community at SLAC.

Force balance operation was achieved with an electrostatic actuator, while a second piezoelectric driven actuator was used to compensate for the creep of the spring system (see Fig.4 and 5). Although sub-nanometer resolution at 1 Hz was proven in preliminary tests, the investigation was stopped after the choice of cold technology for the ILC.

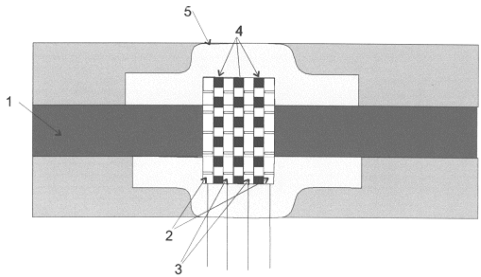

**Figure 6.** Standard MET cell: 1. electrolytic fluid, 2. anodes, 3. cathodes, 4. dielectric spacers, 5. ceramic tube.

A valid alternative solution, available off-the-shelf, are electrochemical seismic sensors, which are based on the so-called molecular electronic transducer (MET) principle [17]. The standard MET cell (see Fig.6) consist of four thin platinum mesh electrodes sandwiched between thin microporous dielectric spacers and placed in a ceramic tube filled with an iodine-based electrolyte. The outer electrodes act as anodes while the inner ones as cathodes. Anodes are connected to a continuous voltage source, while the two cathodes terminals are connected to a high input impedance amplifier. When the voltage source is connected, after a settling time, a quiescent electric current starts flowing between the two pairs of electrodes.

Acceleration causes a pressure differential across the cell which causes a flow of electrolyte between the electrodes. The flow transports ions, raising their concentration on one set of electrodes and reducing it on the other. This produces a net change of the electric current in the readout circuit proportional to the acceleration of the cell. Minimum detectable inertial displacements better than $0.1\,\text{nm}/\sqrt{\text{Hz}}$ at 1Hz are achievable, limited by Brownian fluctuations of the flow, due to the hydrodynamic impedance of the cell (analogue to the Johnson noise of a resistor). Recently, a modified version of this sensor, immune to very strong magnetic fields (up to 6 T) and highly radiation hardened, has been developed by PMD Scientific, Inc. [18] in collaboration with SLAC investigators and it is, at present, the most promising way to have a witness inertial sensor aboard the FD's cryostats. Many concerns, now going to be addressed experimentally, have been made about vibrations of the FD superconducting coils inside their cryostats. For this reason, interferometric position sensors, looking at the cold mass from the cryostat wall or from another external reference (better if common for both sides of machine, as in the well known optical anchor idea), are being developed [19]. A state-of-the-art commercial laser velocimeter, based on a heterodyne Mach-Zender interferometer and capable of measuring differential displacements at one nanometer level, is currently used at Brookhaven National Laboratory to study superconducting magnet vibrations produced by the liquid helium cooling system [20]. Possible on-board active stabilization techniques are on the way [21]. Use of non-contacting position sensors fitting in the cryostat could also be evaluated and viable. The most suitable choice could be core-less LVDT's (linear variable differential transformers). These sensors, widely used in gravitational wave community for low frequency controls [22], can operate in the very prohibitive IP environment, even at liquid helium temperature, and provide nanometer level resolution, with very high linearity over one-centimeter scale dynamic range. In case of severe space limitations, planar design is also implementable without loss of resolution.

## 4. Conclusions

Mechanical vibration diagnostics will play a relevant role for the ILC, both for the commissioning of the machine and for the optimization of the beam-based feedback systems. The required resolution for the sensors, lying in the range of nanometers, is properly scaled for the tolerances accepted for the operation of the machine. In linac sections, small size inertial sensors mounted aboard the quadrupoles and capable to operate at liquid helium temperature, seem to be the best solution to monitor uncorrelated vibrations of the focussing optical elements. Displacement sensors, measuring the position of the quadrupole with respect to the main cryostat wall, have been proposed but they are not sufficient for the purpose since, as confirmed by recent experiments, quadrupole vibration spectrum will likely be dominated by the low frequency rigid body motion of the cryomodule vacuum vessel on its support system.

Off the shelf piezoelectric accelerometers, at present the only evaluated inertial sensors, are not reliable at frequencies below 10 Hz, and alternative solutions are being investigated.

For FF section, compact seismometers, based on molecular electronics transducers and able to withstand with strong magnetic fields (up to 6 T), are now commercially available and can be used as witness sensors aboard the FD cryostats inside the detector's yoke. Monitoring of the relative coordinates of the two FD's at the IP with interferometric position sensors is the preferred choice foreseen in the BCD, and suitable instrumentation is being designed, even if the feasibility in the crowded and prohibitive environment of the detector area has not proven yet.

## Acknowledgements

The author would like to acknowledge members and collaborators of the DESY Ground Vibration group, Wilhelm Bialowons and Eckhard Elsen for useful comments. This work is supported by the Commission of the European Communities under the 6[th] Framework Programme "Structuring the European Research Area", contract number RIDS-011899.